\documentclass[12pt]{article}
\pdfoutput=1

\usepackage{amsmath,amsfonts,amssymb,epsfig}
\usepackage{slashed}
\numberwithin{equation}{section}

\usepackage{graphicx}
\usepackage{cancel}
\usepackage{cite}
\usepackage{color}

\usepackage{xspace}

\def\pdir{Paperplots/}

\newcommand{\exclude}[1]{}
\def\nn{\nonumber}

\def\beq{\begin{equation}}
\def\eeq{\end{equation}}
\def\bal{\begin{align}}
\def\eal{\end{align}}

\def\s2b{s_{2\beta}}
\def\c2b{c_{2\beta}}

\def\2b2[#1,#2][#3,#4]{\left( \begin{array}{cc} #1 & #2 \\ #3 & #4 \end{array}
\right)}
\def\3b3[#1,#2,#3][#4,#5,#6][#7,#8,#9]{\left( \begin{array}{ccc} #1 & #2 &#3 \\
#4 & #5 & #6\\#7&#8&#9\end{array} \right)}
\def\thv[#1,#2,#3]{\left( \begin{array}{c} #1 \\ #2 \\ #3 \end{array} \right)}
\def\twv[#1,#2]{\left( \begin{array}{c} #1 \\ #2 \end{array} \right)}

\def\ov{\overline}

\def\smallSM{{\rm{\scriptscriptstyle SM}}}
\def\smallS{{\scriptscriptstyle S}}
\def\smallZ{{\scriptscriptstyle Z}}
\def\smallW{{\scriptscriptstyle W}}
\def\smallB{{\scriptscriptstyle B}}
\def\smallH{{\scriptscriptstyle H}}
\def\smallF{{\scriptscriptstyle F}}
\def\MS{M_\smallS}
\def\MZ{M_\smallZ}
\def\MW{M_\smallW}
\def\MH{M_\smallH}
\def\UF{U(1)_\smallF}
\def\MGUT{M_{\rm {\scriptscriptstyle GUT}}}
\def\tBp{{\tilde \smallB^\prime}}
\def\tWp{{\tilde \smallW^\prime}}
\def\tgp{{\tilde g^\prime}}
\def\mtBp{m_\tBp}
\def\mtWp{m_\tWp}
\def\mtgp{m_\tgp}
\def\mtB{m_{\tilde \smallB}}
\def\mtW{m_{\tilde \smallW}}

\def\Mtgp{M_\tBp}

\def\Mtgp{M_\tgp}
\def\msbar{{\ov {\rm MS}}}
\def\beq{\begin{equation}}
\def\eeq{\end{equation}}
\def\bea{\begin{eqnarray}}
\def\eea{\end{eqnarray}}

\newenvironment{Appendix}
 {
  \setcounter{section}{0}
  \setcounter{equation}{0}
  
 }

\begin{document}

\begin{flushright}
\end{flushright}
\begin{center}

\vspace{1cm}
{\LARGE{\bf A Fake Split-Supersymmetry Model  \\[3mm]
for the 126 GeV Higgs}}

\vspace{1cm}

\large{  Karim Benakli$^\clubsuit$ \let\thefootnote\relax\footnote{$^\clubsuit$kbenakli@lpthe.jussieu.fr},
Luc Darm\' e$^\heartsuit$ \footnote{$^\heartsuit$darme@lpthe.jussieu.fr},
Mark~D.~Goodsell$^\diamondsuit$ \footnote{$^\diamondsuit$goodsell@lpthe.jussieu.fr}
and  Pietro Slavich$^\spadesuit$  \footnote{$^\spadesuit$slavich@lpthe.jussieu.fr} \\[5mm]}

{\small
\emph{1-- Sorbonne Universit\'es, UPMC Univ Paris 06, UMR 7589, LPTHE, F-75005, Paris, France \\
2-- CNRS, UMR 7589, LPTHE, F-75005, Paris, France }}

\end{center}
\vspace{0.7cm}

\abstract{We consider a scenario where supersymmetry is broken at a
  high energy scale, out of reach of the LHC, but leaves a few
  fermionic states at the TeV scale.  The particle content of the
  low-energy effective theory is similar to that of Split
  Supersymmetry. However, the gauginos and higgsinos are replaced by
  fermions carrying the same quantum numbers but having different
  couplings, which we call fake gauginos and fake higgsinos.  We study
  the prediction for the light-Higgs mass in this Fake Split-SUSY
  Model (FSSM).  We find that, in contrast to Split or High-Scale
  Supersymmetry, a 126 GeV Higgs boson is easily obtained even for
  arbitrarily high values of the supersymmetry scale
  $\MS$. For $\MS \gtrsim 10^8$ GeV, the Higgs mass is almost
  independent of the supersymmetry scale and the stop mixing
  parameter, while the observed value is achieved for $\tan{\beta}$
  between $1.3$ and $1.8$ depending on the gluino mass. }

\newpage

\tableofcontents

\setcounter{footnote}{0}

\section{Introduction}

The LHC experiments have completed the discovery of all of the
particles predicted by the Standard Model (SM). The uncovering of the
last building block, the Higgs boson \cite{Atlas:2012gk,CMS:2012gu},
opens the way for a more precise experimental investigation of the
electroweak sector. Of particular interest is understanding the
possible role of supersymmetry.

Supersymmetry (SUSY) can find diverse motivations. From a lower-energy
point of view, (i) it eases the problem of the hierarchy of the
gauge-symmetry-breaking scale versus the Planck scale; (ii) it
provides candidates for dark matter; (iii) it allows unification of
gauge couplings and even predicts it within the Minimal Supersymmetric
Standard Model (MSSM).  On the other hand, supersymmetry can be
motivated as an essential ingredient of the ultraviolet (UV) theory,
having String Theory in mind.  In the latter framework, there is no
obvious reason to expect supersymmetry to be broken at a particular
scale, which is usually requested to be much below the fundamental
one. The original motivations of low-energy supersymmetry might then
be questioned. In fact, the Split-Supersymmetry
model~\cite{ArkaniHamed:2004fb, Giudice:2004tc, ArkaniHamed:2004yi}
abandons (i) among the motivations of supersymmetry, while retaining
(ii) and (iii).  The idea of Split SUSY is to consider an MSSM content
with a split spectrum. All scalars but the lightest Higgs are taken to
be very massive, well above any energy accessible at near-future
colliders, while the gauginos and the higgsinos remain light, with
masses protected by an approximate $R$-symmetry.

It is important to emphasise that, even if supersymmetry is broken at
an arbitrarily high scale $\MS$, its presence still has implications
at low energy for the Higgs mass. Indeed, in supersymmetric models the
value of the Higgs quartic coupling is fixed once the model content
and superpotential couplings are given.  This provides a boundary
condition at $\MS$ for the renormalisation group (RG) evolution down
to the weak scale to predict the value of the Higgs mass.  As a
result, in the Split-SUSY model the prediction for the light-Higgs
mass can be in agreement with the measured value of about $126$ GeV
only for values of $\MS$ not exceeding about $10^9$ GeV (see for
example refs~\cite{Bernal:2007uv, Giudice:2011cg}).

In this paper we consider a new scenario.  In the spirit of Split
SUSY, we assume that fine-tuning is responsible for the presence of a
light Higgs. A first difference is that our UV model is not the MSSM
but it is extended by additional states in the adjoint representation
of the SM gauge group, as in ref.~\cite{Fayet:1978qc}. Such a field
content has been discussed in the so-called Split Extended
SUSY~\cite{Antoniadis:2005em, Antoniadis:2006eb} (see also
\cite{Carena:2004ha,Unwin:2012fj} for related work), where it was
assumed that the additional gaugino-like and higgsino-like
states arise as partners of the SM gauge bosons under an extended
supersymmetry, and different hierarchies between the Dirac and
Majorana masses have been considered in ref.~\cite{Belanger:2009wf}. Furthermore, 
in related work a similar scenario to ours was recently presented in \cite{Dudas:2013gga}.

A fundamental difference between our scenario and the usual Split SUSY
or the closely related models mentioned above is that {$R$-symmetry is
  strongly broken} and does not protect the gauginos from obtaining
masses comparable to the scalar ones \cite{Dudas:2013gga}. 
In the simplest realisation presented here, in order to keep the extra states light, we endow them
with charges under a new $\UF$ symmetry. An $N=2$ supersymmetry origin
of the new states~\cite{Antoniadis:2005em, Antoniadis:2006eb} raises
then the difficulty of embedding $\UF$ in an $R$-symmetry and will not
be discussed here.

For $\MS$ lower than the MSSM GUT scale $\MGUT \approx 2\!\times\!
10^{16}$ GeV, the achievement of unification requires additional
superfields which restore convergence of the three SM gauge
couplings. This set of extra states can be chosen to be the ones that
are required for unification of gauge couplings in Dirac gaugino
unified models~\cite{MDGSSM}, and can safely be assumed to
appear only above $\MS$, not affecting the discussion in this work:
the properties of our model are fixed at $\MS$ and, as we shall
establish, any corrections that we cannot determine are tiny.

Below the supersymmetry  scale $\MS$, the field content of the
model is the same as in the usual Split SUSY, but the gauginos are
replaced by very weakly coupled fermions in the adjoint representation
that we call ``fake gauginos'', and the higgsinos are replaced by
weakly coupled fermion doublets that we call ``fake higgsinos''.  At
the TeV scale the model looks like Split SUSY with fake gauginos and
higgsinos, hence the name of Fake Split-SUSY Model (FSSM). As we will
show, a remarkable consequence of the different couplings of the fake
gauginos and higgsinos to the Higgs boson, compared to the usual
gauginos and higgsinos, is that a prediction for the Higgs mass
compatible with the observed value can be obtained for arbitrarily
high values of $\MS$.

The plan of the paper is as follows. In section \ref{SEC:FSSM}, we
describe the field content of the Fake Split-SUSY Model and a possible
realisation using a broken additional $\UF$ symmetry.  The latter will
be at the origin of the desired hierarchy between different couplings
and mass parameters. We explain how the effective field theory of the
FSSM compares with the usual Split SUSY. Section \ref{SEC:Constraints}
briefly discusses the collider and cosmological constraints. Section
\ref{SEC:Higgs} presents the predictions of the model for the Higgs
mass. The assumptions, inputs and approximations used in the
computation are described in section \ref{sec:procedure}, while
numerical results are presented in section \ref{sec:results}.  We also
provide a comparison with the cases of Split SUSY and High-Scale SUSY,
showing the improvement for fitting the experimental value of the
Higgs mass for arbitrarily high values of the supersymmetry scale
$\MS$. Our main results, and open questions requiring further
investigation, are summarised in the conclusions. Finally, the
two-loop renormalisation group equations (RGEs) for the mass parameters
of Split SUSY are given in an appendix.
 
\section{The Fake Split-Supersymmetry Model}
\label{SEC:FSSM}

\subsection{The model at the SUSY scale}
\label{SEC:susymodel}

At the high SUSY scale $\MS$, we extend the MSSM by additional chiral
superfields and a $\UF$ symmetry. There are three sets of additional
states:~\footnote{In the following, bold-face symbols denote
  superfields.}

\begin{enumerate}

\item \emph{Fake gauginos} (henceforth, F-gauginos) are fermions
  $\chi_\Sigma$ in the adjoint representation of each gauge group,
  which sit in a chiral multiplet $\mathbf{\Sigma}$ having scalar
  component $\Sigma$. These consist of: a singlet $\mathbf{S} = S +
  \sqrt{2} \theta \chi_{S }+ \ldots$ ; an $SU(2)$ triplet $\mathbf{T}
  = \sum_a \mathbf{T}^{a} \,\sigma^a/2$, where $ \mathbf{T}^{a} =
  T^{a} + \sqrt{2} \theta \chi_T^{a} + \ldots $ and $\sigma^a$ are the
  three Pauli matrices; an $SU(3)$ octet $\mathbf{O} = \sum_a
  \mathbf{O}^a\, \lambda^a/2$, where $\mathbf{O}^a = O^{a} + \sqrt{2}
  \theta \chi_O^{a}+ \ldots$ and $\lambda^a$ are the eight Gell-Mann
  matrices.

\item Higgs-like $SU(2)_W$ doublets $\mathbf{H_u^\prime}$ and
  $\mathbf{H_d^\prime}$ (henceforth, F-Higgs doublets) with fermions
  appearing as \emph{fake higgsinos} (henceforth, F-higgsinos).

\item Two pairs of vector-like electron superfields (i.e.~two pairs of
  superfields with charges $\pm 1$ under $U(1)_Y$) with a
  supersymmetric mass $\MS$. For $\MS \lesssim \MGUT$ these fields
  restore the possibility of gauge coupling unification, because they
  equalise the shifts in the one-loop beta functions at $\MS$ of all
  of the gauge groups relative to the MSSM~\cite{MDGSSM}.

\end{enumerate}

In contrast to the usual Split-SUSY case -- and also in contrast to
the usual Dirac gaugino case -- we do not preserve an
$R$-symmetry. This means that the gauginos have masses at $\MS$,
moreover the higgsino mass is not protected, thus a $\mu$ term of
order $\MS$ will be generated for the higgsinos.

However, we introduce an approximate $\UF$ symmetry under which all
the adjoint superfields and the F-Higgs fields $\mathbf{H_u^\prime}$
and $\mathbf{H_d^\prime}$ have the same charge. The breaking of this
symmetry is determined by a small parameter $\varepsilon$ which may
correspond to the expectation value of some charged field divided by
the fundamental mass scale of the theory (at which Yukawa couplings
are generated); this reasoning is familiar from flavour models. We can
write the superpotential of the Higgs sector of the theory
schematically as
\bea
\label{Superpotential}
W &=&  \mu_0 \, \mathbf{H_u\cdot H_d } 
+ Y_u \, \mathbf{U^c \, Q \cdot H_u} 
- Y_d \, \mathbf{D^c \, Q \cdot H_d} 
- Y_e \, \mathbf{E^c \, L \cdot H_d} \nn\\
& +& \varepsilon \,\left(
\hat{\mu}'_d \,\mathbf{H_u\cdot H_d^\prime } + 
\hat{\mu}'_u \,\mathbf{H_u^\prime \cdot H_d }
+ \hat{Y}_u^\prime\, \mathbf{U^c\,Q \cdot H_u^\prime} 
- \hat{Y}_d^\prime\, \mathbf{D^c\,Q \cdot H_d^\prime} 
- \hat{Y}_e^\prime\, \mathbf{E^c\,L \cdot H_d^\prime}\right)
\nn\\
& +&  \varepsilon\, \left(
\hat{\lambda}_S \, \mathbf{S \, H_u\cdot H_d}  
+ 2\, \hat{\lambda}_T\, \mathbf{H_d\!\cdot\! T\, H_u}\right) 
\nn\\
& +& \varepsilon^2 \,\left(
\hat{\lambda}_{Sd}^\prime\, \mathbf{S \,H_u\cdot H_d^\prime }  
+ \hat{\lambda}_{Su}'\, \mathbf{S \,H_u^\prime\cdot H_d } 
+ 2\, \hat{\lambda}_{Tu}^\prime \, \mathbf{H_d\!\cdot\! T \,H_u^\prime} 
+ 2\, \hat{\lambda}_{Td}^\prime\, \mathbf{H_d^\prime\!\cdot\! T \,H_u}
\right)
\nn\\
& +& 
\varepsilon^2\,\hat{\mu}''\, \mathbf{H_u^\prime \cdot H_d^\prime }
~+~
\varepsilon^2 \,\left[
\frac12 \,\hat{M}_S \,\mathbf{S}^2 + \hat{M}_T\,
\textrm{Tr}(\mathbf{TT}) + \hat{M}_O \,\textrm{Tr}(\mathbf{OO}) \right]~,
\eea
where we have neglected irrelevantly small terms of higher order in
$\varepsilon$.  Even if chosen to vanish in the supersymmetric theory,
some parameters in eq.~(\ref{Superpotential}), such as the bilinear
$\mu$ terms, obtain contributions when supersymmetry is broken. In
order to keep track of the order of suppression, we have explicitly
extracted the parametric dependence on $\varepsilon$ due to the
$U(1)_F$ charges,\footnote{We use a $\widehat{\mathrm{hat}}$ to denote
  the suppressed terms.}  such that all the mass parameters are of
${\cal O}(\MS)$, and all the dimensionless couplings are either of
order one or suppressed by loop factors.

Note that the ``fake states'' can appear as partners of the MSSM gauge
bosons under an extended $N=2$ supersymmetry that is \emph{explicitly}
broken at the UV scale to $N=1$. The imprint of $N=2$ is the extension
of the states in the gauge sector into gauge vector multiplets and
Higgs hyper-multiplets which give rise to the fake gauginos and
higgsinos when broken down to $N=1$.  The quarks and leptons of the
MSSM should be identified with purely $N=1$ states. The difficulty of
such a scenario resides in making only parts of the $N=2$ multiplets
charged under $\UF$. It is then tempting to identify the $\UF$ as 
part of the original $R$-symmetry. We will not pursue the discussion
of such possibility here.

We will now review the spectrum of states resulting from
eq.~(\ref{Superpotential}).
The Higgs soft terms, and thence the Higgs mass matrix, can be written
as a matrix in terms of the four-vector $ v_H\equiv (H_u, {H_d}^*,
H_u^\prime, H_d^{\prime\,*})$
\begin{align}
-\frac{1}{\MS^2}\mathcal{L}_{soft} ~~\supset ~~& v_H^\dagger \left(
\begin{array}{cccc} \mathcal{O}(1) & \mathcal{O}(1) & \mathcal{O}(\varepsilon) &
\mathcal{O}(\varepsilon) \\ \mathcal{O}(1) &
\mathcal{O}(1) & \mathcal{O}(\varepsilon) &
\mathcal{O}(\varepsilon) \\ \mathcal{O}(\varepsilon) & \mathcal{O}(\varepsilon)
& \mathcal{O}(1) & \mathcal{O}(\varepsilon^2) \\
\mathcal{O}(\varepsilon) &
\mathcal{O}(\varepsilon)& \mathcal{O}(\varepsilon^2) & \mathcal{O}(1)
\end{array} \right) v_H .
\end{align}
In the spirit of the Split-SUSY scenario, the weak scale is tuned to
have its correct value, and the SM-like Higgs boson is a linear
combination of the original Higgs and F-Higgs doublets: 
\beq
\label{eq:lincomb1}
H_u ~\approx~ \, \sin\beta\, H\, + \ldots~, 
~~~ \qquad H_d ~\approx~ \,
\cos\beta\,i\sigma^2\, H^* \,+ \dots~,
\eeq
\beq
\label{eq:lincomb2}
H_u^\prime ~\approx~ \varepsilon\, H \,+ \ldots~,~~~
 \qquad H_d^\prime ~ \approx \varepsilon \,
i\sigma^2\,H^* \,+ \ldots~,
\eeq
where $\beta$ is a mixing angle and the ellipses stand for terms of
higher order in $\varepsilon$. Due to the suppression of the mixing
between the eigenstates by the $\UF$ symmetry, this pattern is
ensured. Note that, if we wanted to simplify the model, we could
impose an additional unbroken symmetry under which the F-Higgs fields
transform and are vector-like -- for example, lepton number. In this
way we would remove the mixing between the Higgs and F-Higgs
fields. This is unimportant in what follows, since we are only
interested in the light fields that remain.

Eqs~(\ref{eq:lincomb1}) and (\ref{eq:lincomb2}) show that the SM-like
Higgs boson is, to leading order in $\varepsilon$, a linear
combination of the fields $H_u$ and $H_d$.  Thus, the Yukawa couplings
are unaffected compared to the usual Split-SUSY scenario. The original
higgsinos are rendered heavy, while the light fermionic eigenstates
consist of $\tilde{H}_u^\prime$ and $\tilde{H}_d^\prime\,$, with mass
$\mu$ of ${\cal O}(\varepsilon^2 \MS)$ and an ${\cal O}(\varepsilon)$
mixing with the original higgsinos.

Since we are not preserving an $R$-symmetry, the original gaugino
degrees of freedom will obtain masses of ${\cal O}(\MS)$, and we will
also generate $A$-terms of the same order (although there may be some
hierarchy between them if supersymmetry breaking is gauge-mediated). 
On the other hand, since the adjoint fields transform under
a (broken) $\UF$ symmetry, Dirac mass terms for the gauginos and also
masses for the adjoint fermions are generated by supersymmetry
breaking, but they are suppressed by one and two powers of
$\varepsilon$, respectively.
We can write the masses for the gauginos $\lambda$ and the
adjoint fermions $\chi$ as 
\beq
 -\Delta\mathcal{L}_{\rm gauginos} = \MS \,\bigg[  
\frac{1}{2}\, \lambda \lambda 
~+~ \mathcal{O}(\varepsilon)\,  \lambda \chi 
~+~ \mathcal{O}(\varepsilon^2)\chi  \chi 
~+~{\rm h.c.} ~\bigg]~,
\eeq
giving a gaugino/F-gaugino mass matrix
\beq
\mathcal{M}_{1/2}~\sim~ \MS\, \begin{pmatrix}
1    &   \mathcal{O}(\varepsilon)  \\
  \mathcal{O}(\varepsilon) &\mathcal{O}(\varepsilon^2)
\end{pmatrix}.
\eeq
This leaves a heavy eigenstate of ${\cal O}(\MS)$ and a light one of
${\cal O}(\varepsilon^2 \MS)$, where the light eigenstate is to
leading order $\chi + \mathcal{O}(\varepsilon)\,\lambda$.

We will assume that the Dirac masses are generated by D-terms of
similar order to the $R$-symmetry-violating F-terms. This means that
$B$-type mass terms for the adjoint scalars are generated of ${\cal
  O}(\varepsilon^2 \MS^2)$ too. However, the usual
supersymmetry-breaking masses for the adjoint scalars $S, T, O$ will
not be suppressed, and therefore will be at the scale $\MS$:
\beq
-\Delta\mathcal{L}_{\rm adjoint\ scalars} = 
\MS^2 \,\bigg[ |\Sigma|^2 +
\mathcal{O}(\varepsilon^2) 
( \frac{1}{2} \Sigma^2 + \frac{1}{2} \Sigma^{*\,2} )
\bigg]~.
\eeq
This is straightforward to see in the case of gravity mediation, and
in the case of gauge mediation we see that the triplet/octet adjoint
scalars acquire these masses -- as the sfermions do -- at two loops
(while in this case the singlet scalar would have a mass at an
intermediate scale, but couplings to all light fields
suppressed). This resolves in a very straightforward way the problem,
typical of Dirac gaugino models, of having tachyonic adjoints
\cite{Benakli:2008pg,Benakli:2010gi,Csaki:2013fla}.

\subsection{Below \boldmath{$\MS$}, the FSSM}
\label{sec:fssm}

Below the supersymmetry scale $\MS$, we can integrate out all
of the heavy states and find that the particle content of the theory
appears exactly the same as in Split SUSY: this is why we call the
scenario Fake Split SUSY. Above the electroweak scale, we have F-Binos
$\tilde{B}^\prime$, F-Winos $\tilde{W}^\prime$ and F-gluinos
$\tilde{g}^\prime$ with (Majorana) masses $\mtBp$, $\mtWp$ and
$\mtgp$, respectively, and F-higgsinos $\tilde{H}_{u,d}^\prime $ with
a Dirac mass $\mu$.

We can also determine the effective renormalisable couplings.  The
F-gauginos and F-higgsinos have their usual couplings to the gauge
fields.  The F-gluinos have only gauge interactions, whereas there are
in principle renormalisable interactions between the Higgs,
F-higgsinos and F-electroweakinos. The allowed interactions take the
form
\begin{align}
\label{eq:gtildas}
  & {\cal L}_{\rm eff} \supset -\frac{H^{\dagger}}{\sqrt{2}}
  (\tilde{g}_{2u}\, \sigma^a\, \tilde {W^\prime}^a + \tilde{g}_{1u}\, \tilde
  {B}^\prime) \ \tilde {H}_u^\prime - \frac{H^{T} i \sigma^2}{\sqrt{2}}
  (- \tilde {g}_{2d} \,\sigma^a\, \tilde {W^\prime}^a + \tilde{g}_{1d}\,
  \tilde{B}^\prime) \ \tilde{H}_d^\prime~.
\end{align}
Since the gauge couplings of all the particles are the same as in the
usual Split-SUSY case, the allowed couplings take the same
form. However, the values differ greatly. The couplings in
eq.~(\ref{eq:gtildas}) descend from the gauge current terms, given by
\begin{align}
  {\cal L}_{\rm gauge\ current} \supset& -
  \frac{H^{\dagger}_u}{\sqrt{2}} (g\, \sigma^a \lambda_2^a + g^\prime\,
  \lambda_Y) \ \tilde {H}_u - \frac{H_d^\dagger}{\sqrt{2}} ( g\,
  \sigma^a\lambda_2^a - g^\prime\, \lambda_Y)
  \ \tilde{H}_d \nn\\
  &- \frac{H^{\prime\, \dagger}_u}{\sqrt{2}} (g \,\sigma^a \lambda_2^a +
  g^\prime\, \lambda_Y) \ \tilde {H}^\prime_u -
  \frac{H_d^{\prime\, \dagger}}{\sqrt{2}} ( g\, \sigma^a\lambda_2^a -
  g^\prime\, \lambda_Y) \ \tilde{H}_d^\prime~,
\end{align}
where $\lambda_2, \lambda_Y $ are the gauginos of $SU(2)$ and
hypercharge in the high-energy theory, but there are also terms of the
same form from the superpotential terms $\varepsilon
\hat{\lambda}_{S,T}$, $\varepsilon^2 \hat{\lambda}_{Su,d}^\prime$,
$\varepsilon^2 \hat{\lambda}_{Tu,d}^\prime $ involving the fields
$\chi_S$ and $\chi_T$. When we integrate out the heavy fields, we then
see that in our model the couplings are \emph{doubly} suppressed:
\begin{equation}
\tilde{g}_{1u}  \sim \tilde{g}_{1d}  \sim \tilde{g}_{2u}  
\sim \tilde{g}_{2d}  \sim \varepsilon^2.
\end{equation}
We recall that, in the usual Split-SUSY case, we would have instead
$\tilde{g}_{2u} = g \sin \beta$, $\tilde{g}_{2d} = g \cos \beta$,
$\tilde{g}_{1u} = g^\prime \sin \beta$ and $\tilde{g}_{1d} = g^\prime
\cos \beta$, where $\beta$ is the angle that rotates the Higgs
doublets $H_u$ and $H_d$ into one light, SM-like doublet and a heavy
one.

The remaining renormalisable coupling in the theory is the Higgs
quartic coupling $\lambda$, which at tree level is determined by
supersymmetry to be
\beq
\label{lambda0}
\lambda ~=~
\frac14\left(g^2+g^{\prime\,2}\right)\,\cos^22\beta ~+~{\cal
  O}(\varepsilon^2)~. 
\eeq
The tree-level corrections at $\mathcal{O}(\varepsilon^2)$ come from the superpotential 
couplings $\hat{\lambda}_S$ and $\hat{\lambda}_T$, and from the $\mathcal{O}(\varepsilon)$ mixing between the Higgs and F-Higgs fields.  Additional $\mathcal{O}(1)$ contributions to this relation could arise if the SUSY model above  $\MS$ included new, substantial superpotential (or D-term) interactions involving the SM-like Higgs, but this is not the case for the model described in section \ref{SEC:susymodel}. There are, however, small loop-level corrections to eq.~(\ref{lambda0}), which we will discuss in section \ref{SEC:Higgs}. 

The $\mathcal{O}(\varepsilon^2)$ corrections to the
$\tilde{g}_{(1,2)(u,d)}$ and $\lambda$ couplings are not determined
from the low-energy theory and are thus unknown. However, in this
study we focus on models where the set of F-gauginos and F-higgsinos
lies in the TeV mass range, which corresponds to values of
$\varepsilon$ of the order of
\begin{equation}
\varepsilon \sim \sqrt{\frac{{\rm TeV}}{\MS} }~,
\end{equation}
which gives a $\varepsilon^2$ ranging between $10^{-13}$ to $10^{-2}$
when $\MS$ goes from the highest GUT scale of $10^{16}$ GeV down to
$100$ TeV, the lowest scale considered here. With such values of
$\varepsilon$, we have verified that we can safely neglect the
contribution of $\tilde{g}_{(1,2)(u,d)}$ to the running of the Higgs
quartic coupling, and that the shift in the Higgs mass due to the
tree-level corrections to $\lambda$ is less than $2$ GeV for $\MS >
100$ TeV, falling to a negligibly small amount for $\MS > 1000$ TeV.

\subsubsection{Gauge coupling unification}

One of the main features of the MSSM that is preserved by the split
limit is the unification of gauge couplings. At the one-loop level,
the running of gauge couplings in our model is the same as in
Split SUSY because the Yukawa couplings only enter at two-loop
level. However, we have verified that gauge-coupling unification is
maintained at two loops in our model.

\subsubsection{Mass matrices}

From the discussion above we can then read off the mass matrices after
electroweak symmetry breaking. In the basis
$(\tilde{B}^\prime,\tilde{W}^{\prime\,0}, \tilde{H'_d}^0,
\tilde{H'_u}^0)$ the neutralino mass matrix is~\footnote{From now on,
  given the smallness of $\varepsilon$, we shall not keep explicit
  track of the numerical coefficients in front of it, thus we will use
  $\varepsilon^n$ as a shorthand for $\mathcal{O}(\varepsilon^n)$. }
\begin{equation}
\mathcal{M}_{\chi^0} = \left(\begin{array}{c c c c}
 \mtBp   & 0       &   \varepsilon^2 \MZ & \varepsilon^2 \MZ
\\
0     & \mtWp  &     \varepsilon^2 \MZ &   \varepsilon^2 \MZ \\
 \varepsilon^2 \MZ &  \varepsilon^2 \MZ  & 0  & -\mu \\
 \varepsilon^2 \MZ  &   \varepsilon^2 \MZ & -\mu  & 0 \\
\end{array}\right)~.
\label{diracgauginos_NeutralinoMassarray2}
\end{equation}
 We see that there is a mixing suppressed by
$\varepsilon^2 = \frac{\mathrm{TeV}}{\MS}$. For example, if the
F-higgsino is the lightest eigenstate, it will be approximately Dirac
with a splitting of the eigenvalues of order $\varepsilon^4 \MZ^2/\mu
\sim \left(\frac{ \mathrm{TeV}}{\MS}\right)^2 \MZ $.

We then write the chargino mass matrix involving the $\tilde{H'}^+$,
$\tilde{H'}^-$ and the charged F-gauginos $\tilde{W^\prime}^+$ and
$\tilde{W^\prime}^-$. The mass terms for the charginos can be expressed in
the form
\begin{equation}
- (v^-)^T\mathcal{M}_{\chi^\pm}  v^+ ~+~ {\rm h.c.}~,
\label{diracgauginos_CharginoMassLagrangian}
\end{equation}
where we have adopted the basis $v^+ =
(\tilde{W^\prime}^+,\tilde{H'}^+_u)$, $v^- =
(\tilde{W^\prime}^-,\tilde{H'}^-_d)$. This gives
\begin{equation}
\mathcal{M}_{\chi^\pm}~ =~
\left(\begin{array}{c c}
\mtWp  &  \varepsilon^2 \MW \\
 \varepsilon^2 \MW  & \mu \\
\end{array}\right) .
\label{diracgauginos_CharginoMassarray2}
\end{equation}
Again we have very little mixing.

Clearly, the mixing coefficients of order $\varepsilon^2$ in the mass
matrices are dependent on quantities in the high-energy theory that we
cannot determine. However, because they are so small, they have
essentially no bearing on the mass spectrum of the theory (although
they will be relevant for the lifetimes).

\section{Comments on cosmology and colliders}
\label{SEC:Constraints}

The signatures of Fake Split SUSY concern the phenomenology of the
F-higgsinos and F-gauginos, and thus share many features with the
usual Split-SUSY case. They differ quantitatively in that the
lifetimes are parametrically enhanced: the decay of heavy neutralinos
and the F-gluino to the lightest neutralino must all proceed either
via $\varepsilon^2$-suppressed mixing terms or via sfermion
interactions, and, since the F-higgsinos/gauginos only couple to
sfermions via mixing, each vertex is therefore suppressed by a factor
of $\varepsilon$ or $\varepsilon^2$. Hence the lifetimes are enhanced
by a factor of $\varepsilon^{-4}$ \cite{Gambino:2005eh,Dudas:2013gga};
in particular the F-gluino lifetime is
\begin{align}
 \tau_{\tilde g'} ~\simeq~&  \frac{4 \ \text{sec}}{\varepsilon^{4}}
\times\left(\frac{M_S}{10^9\text{GeV}}\right)^4
\times\left(\frac{1 \ \text{TeV}}{\mtgp}\right)^5 \nn\\
  ~\sim~&
  \text{sec}\times\left(\frac{M_S}{10^7\text{GeV}}\right)^6
\times\left(\frac{1     \ \text{TeV}}{\mtgp}\right)^7~,
\end{align} 
where on the second line we used $\mtgp = \varepsilon^2 \MS $. The
constraints from colliders then depend upon whether the gluino decays
inside or outside the detector; the latter will occur for $M_S \gtrsim
1000$ TeV. In this case, bounds can still be set because the gluino
hadronises and can therefore leave tracks in the detector; the
subsequent R-hadron can collect electric charge that can be detected
in a tracker and/or muon chamber. The bounds on the gluino mass now
reach to about $1.3$ TeV
\cite{Khachatryan:2011ts,Aad:2011yf,Chatrchyan:2013oca,Aad:2013gva}
with the exact bound dependent on the model of hadronisation of the
gluino.

The gluino lifetime is also crucial for determining the cosmology of
the model \cite{ArkaniHamed:2004fb,Arvanitaki:2005fa}. In the standard
Split-SUSY case, if the gluino has a lifetime above $100$ seconds then
it would be excluded when assuming a standard cosmology
\cite{Arvanitaki:2005fa} due to constraints from Big Bang
Nucleosynthesis (BBN). In our case, this would limit $M_S \lesssim
10^7$ GeV. While the bound is no longer necessarily exact, because the
relationship between the mass and lifetime is different in our case,
it still approximately applies. If, on the other hand, the gluino
decays well after the end of BBN such that it deposits very little
energy at BBN times, then other constraints become relevant: it can
distort the CMB spectrum and/or produce photons visible in the diffuse
gamma-ray background. Finally, when the gluino becomes stable compared
to the age of the universe, in our case corresponding to $M_S \gtrsim
10^{10}$ GeV, very strong constraints from heavy-isotope searches
become important, as we shall briefly discuss below.

\subsection{Gravitino LSP}

One way to attempt to allow the gluino to decay is to have a gravitino
LSP. In minimally coupled supergravity, the gravitino has mass
$\frac{F}{\sqrt{3} M_P}$, where $F$ is the order parameter of
supersymmetry breaking. If supersymmetry breaking is mediated at
tree level to the scalars, the supersymmetry scale could be
as high as $\sqrt{F}$ (we could even have some factors of
$\pi$ if we allow for a strongly coupled SUSY-breaking sector, but
that will not substantially affect what follows) and so we could
potentially have a gravitino lighter than the gluino if
\begin{align}
M_S \lesssim&\  5\times 10^{10}\ \mathrm{GeV}\ 
\times\left(\frac{\mtgp}{2\ \mathrm{TeV}} \right)^{1/2}~.
\end{align}
In this case, the F-gluino can decay to a gravitino and either a gluon
or quarks, potentially avoiding the above problems. However, this
relies on the couplings to the goldstino; since we have added Dirac
and fake-gaugino masses, these are no longer the same as in the usual
Split-SUSY case, and a detailed discussion will be given elsewhere
\cite{Dudas:2014inprep}. The effective goldstino couplings are the
Wilson coefficients $C_i^{\tilde{G}}$ of ref.~\cite{Gambino:2005eh},
and in our model we find
\bea
C_i^{\tilde{G}} &=& - \varepsilon \frac{g_s}{\sqrt{2}}~, \qquad i=1...4 \nn\\
C_5^{\tilde{G}} &=&  - \varepsilon \frac{\mtgp }{2\sqrt{2}}~.
\eea
For $i=1...4$ the couplings are to quarks, while the final coupling is
to gluons. We finally obtain the F-gluino width
\beq
\Gamma (\tilde{g}^\prime \rightarrow \tilde{G} + X) ~\simeq ~
 \varepsilon^2\,\frac{\mtgp^5}{2\pi F^2}
\eeq
and hence, for $\MS \sim \sqrt{F}$ (the maximal value), $\mtgp =
\varepsilon^2 \MS$, we find the F-gluino lifetime to be
\beq
\tau_{\tilde{g}^\prime} ~\simeq~ 600\, {\rm sec} 
\times \left(\frac{\MS}{10^6\text{GeV}}\right)^5 
\times\left(\frac{2 \ \text{TeV}}{\mtgp}\right)^6.
\eeq
Hence this cannot be useful to evade the cosmological bounds: the
gravitino couplings are simply too weak.

\subsection{Stable F-gluinos}

For F-gluinos stable on the lifetime of the universe, in our case
corresponding to $\MS \gtrsim 10^{10}$ GeV, remnant F-gluinos could
form bound states with nuclei, which would be detectable as exotic
forms of hydrogen. The relic density is very roughly approximated by
\beq
\Omega_{\tilde{g}} h^2 ~\sim~ \left( \frac{\mtgp}{10\ \mathrm{TeV}}\right)^2,
\eeq
although this assumes that the annihiliations freeze out before the
QCD phase transition and are thus not enhanced by non-perturbative
effects; for heavy F-gluinos this seems reasonable, but in principle
the relic density could be reduced by up to three orders of
magnitude. However, the constraints from heavy-isotope searches are so
severe as to render this moot: the ratio of heavy isotopes to normal
hydrogen $X/H$ should be less than $10^{-29}$ for masses up to $1.2$
TeV \cite{Hemmick:1989ns} or less than $10^{-20}$ for masses up to $10$
TeV \cite{Smith:1982qu}, whereas we find
\beq
\frac{X}{H} ~\sim~ 10^{-4} \left( \frac{\mtgp}{\mathrm{TeV}}\right).
\eeq
If the F-gluino is stable, then we must either:
\begin{enumerate}
\item Dilute the relic abundance of F-gluinos with a late period of
  reheating.
\item Imagine that the reheating temperature after inflation is low
  enough, or that there are several periods of reheating that dilute
  away unwanted relics before the final one.
\end{enumerate}
In both cases, we must ensure that gluinos are not produced during the
reheating process itself, which may prove difficult to arrange: even
if the late-decaying particle decays only to SM fields, if it is
sufficiently massive then high-energy gluons may be among the first
decay products, which could subsequently produce F-gluinos which would
not be able to annihilate or decay away.

The safest solution would be for a decaying scalar to have a mass near
or below twice the F-gluino mass. Then we must make sure that the
decays where the products include only one F-gluino -- and, because of
the residual R-parity, one neutralino -- are sufficiently suppressed,
assuming that the neutralino is somewhat lighter than the
F-gluino. However, such processes are suppressed by a factor of
$\varepsilon^4$, which should sufficiently reduce the branching
fraction of decays by a factor of $10^{20}$ if $\MS \gtrsim 10^{13}$
GeV. Such a scenario would possibly still have difficulty producing
sufficient dilution if the universe is thermal before the final
reheating: suppose that the final reheating occurs when the universe
is at a temperature $T_{\rm decay}$ and reheats the universe to a
temperature $T_R$, then the dilution is of order
$\left(\frac{T_R}{T_{\rm decay}}\right)^3 $. However, if we require
the universe to undergo BBN only once, then both temperatures are
bounded: $T_{\rm decay} > T_{BBN} \sim$ MeV, but also $T_R \lesssim
{\mtgp}/{50}$ to ensure that the \emph{freeze-in} production of
F-gluinos is not too large. Then the amount of dilution achieved is
only of order $10^{14}$ for $2$ TeV F-gluinos, insufficient to evade
bounds from heavy-isotope searches.

We conclude that for a high $\MS \gtrsim 10^{13}$ the most plausible
cosmological scenario is option (2) above: a final reheating
temperature $T_R \lesssim {\mtgp}/{50}$ which occurs either
directly at the end of inflation or after at least one additional
period of low-temperature entropy injection.

\subsection{Neutralinos and dark matter}

Even though the F-gluino may be stable on the lifetime of the universe,
the heavy neutralinos are not (although they may decay on BBN timescales in the 
case of extremely high $\MS$): they can decay to the lightest neutralino
and a Higgs boson via their  $\varepsilon^2$-suppressed Yukawa couplings, so not involving any heavy mass scale. This suppression does however render the 
F-bino effectively inert in the early universe once the heavy neutralinos have decoupled; 
the F-bino would be produced essentially by freeze-in from decays and annihilations of the heavier neutralinos 
-- which have usual weak-scale cross sections and so could potentially thermalise.
Moreover, the charginos will
still decay rapidly via unsuppressed weak interactions to their
corresponding neutralino; the mass splitting between charginos and
neutralinos is produced by loops with electroweak gauge bosons and is
of the order of a few hundred MeV.
If we imagine a modulus in
scenario (2) above that reheats the universe having mass less than
twice that of the F-gluino, but greater than $2 \mtWp$ or $2\mu$, or
where ${\mtWp}/{20}, {\mu}/{20} < T_R \lesssim {\mtgp}/{50}$, we could
potentially have a neutralino dark matter candidate, but the
detailed investigation of this possibility is left for future work.

\section{Fitting the Higgs mass}
\label{SEC:Higgs}

\subsection{Determination of the Higgs mass in the FSSM}
\label{sec:procedure}

Our procedure for the determination of the Higgs-boson mass is based
on the one described in ref.~\cite{Bernal:2007uv} for the regular
Split-SUSY case. We impose boundary conditions on the
$\msbar$-renormalised parameters of the FSSM, some of them at the high
scale $\MS$, where we match our effective theory with the (extended)
MSSM, and some others at the low scale $\MZ$, where we match the
effective theory with the SM. We then use RG evolution iteratively to
obtain all the effective-theory parameters at the weak scale, where we
finally compute the radiatively corrected Higgs mass. However, in
this analysis we improved several aspects of the earlier calculation,
by including the two-loop contributions to the boundary condition for
the top Yukawa coupling, the two-loop contributions to the RG
equations for the Split-SUSY parameters, as well as some two- and
three-loop corrections to the Higgs-boson mass.

At the high scale $\MS$, the boundary condition on the quartic
coupling of the light, SM-like Higgs doublet is determined by
supersymmetry:
\beq
\label{lambdaMS}
\lambda(\MS) ~=~
\frac14\left[g_2^2(\MS)+\frac35\,g_1^2(\MS)\right]\,\cos^22\beta ~+~{\cal
  O}(\varepsilon^2)~, \eeq
where $g_2$ and $g_1$ are the electroweak gauge couplings of the FSSM
in the $SU(5)$ normalisation (i.e.~$g_2=g$ and $g_1 =
\sqrt{5/3}\,g^\prime\,$), $\beta$ is the mixing angle entering
eq.~(\ref{eq:lincomb1}), and the additional terms of ${\cal
  O}(\varepsilon^2)$, which we neglect, arise from suppressed
superpotential couplings and from the mixing of the two MSSM-like
Higgs doublets with the additional F-Higgs doublets.  In contrast with
the Split-SUSY case, a large $\mu_0$-term and $A$-terms  are no longer forbidden by
$R$-symmetry (as the latter is broken at the scale $\MS$), and the
threshold corrections proportional to powers of $|A_t -\mu_0 \cot \beta|^2/\MS^2$ can in
principle alter the boundary condition in eq.~(\ref{lambdaMS}). For
very large values of $\MS$, the top Yukawa coupling that controls
these corrections is suppressed, and their effect on the Higgs mass is
negligible. For lower values of $\MS$, on the other hand, the effect
becomes sizable, and it can shift the Higgs mass by up to 6 GeV when
$\MS \sim 10^5$ GeV~\cite{Giudice:2011cg}. This allows us to obtain
the desired Higgs mass for a lower value of $\tan \beta$ for fixed
$\MS$, or a lower $\MS$ for a given value of $\tan \beta$.  As our
main purpose in this work is to study the possibility of pushing $\MS$
to its highest values, in the following we shall take the stop-mixing
parameter to be vanishing, and we will neglect all of the one-loop
corrections described in refs~\cite{Bernal:2007uv,Giudice:2011cg}.

As mentioned in section \ref{sec:fssm}, the effective
Higgs--higgsino--gaugino couplings $\tilde g_u$, $\tilde g_d$, $\tilde
g^\prime_{u}$ and $ \tilde g^\prime_{d}$ are of ${\cal
  O}(\varepsilon^2)$, and we set them to zero at the matching scale
$\MS$. The RG evolution down to the weak scale does not generate
non-zero values for those couplings, therefore, in contrast with the
case of the regular Split SUSY, the F-higgsinos and F-gauginos have
negligible mixing upon electroweak symmetry breaking, and they do not
participate in the one-loop corrections to the Higgs-boson
mass. Indeed, the electroweak F-gauginos and the F-higgsinos affect
our calculation of the Higgs mass only indirectly, through their
effect on the RG evolution and on the weak-scale boundary conditions
for the electroweak gauge couplings, and we find that the precise
values of their masses have very little impact on the prediction for
the Higgs mass. On the other hand, the choice of the F-gluino mass is
more important due to its effect on the boundary conditions for the
strong and top Yukawa couplings.

To fix the soft SUSY-breaking F-gaugino masses, we take as input the
physical F-gluino mass $\Mtgp$, and convert it to the $\msbar$
parameter $\mtgp$ evaluated at the scale $\Mtgp$ according to the
one-loop relation
\beq
\mtgp(\Mtgp) ~=~ \frac{\Mtgp}{1+\frac{3\,g_3^2}{4\,\pi^2}}~,
\eeq
where $g_3$ is the strong gauge coupling of the FSSM. We then evolve
$\mtgp$ up to the scale $\MS$, where, for simplicity\footnote{Although
  the patterns of neutralino and chargino masses are important for
  collider searches, in our model they have negligible impact on the
  Higgs mass and so the exact relation is not important.}, we impose
on the other two F-gaugino masses the GUT-inspired relations
\beq
\mtBp(\MS)
~=~ \left[\frac{g_1(\MS)}{g_3(\MS)}\right]^2 \,\mtgp(\MS)~,~~~~~~~
\mtWp(\MS)
~=~ \left[\frac{g_2(\MS)}{g_3(\MS)}\right]^2 \,\mtgp(\MS)~.
\eeq
We can then evolve all of the F-gaugino masses down to the
weak scale.  For what concerns the F-higgsino mass $\mu$, we take it
directly as an $\msbar$ input parameter evaluated at the scale $\MZ$.

The gauge and third-family Yukawa couplings, as well as the vacuum
expectation value $v$ of the SM-like Higgs (normalised as $v \approx
174$ GeV), are extracted from the following set of SM
inputs~\cite{Beringer:1900zz,CDF:2013jga}: the strong gauge coupling
$\alpha_s(\MZ) = 0.1184$ (in the $\msbar$ scheme with five active
quarks); the electromagnetic coupling $\alpha(\MZ) = 1/127.944$; the
$Z$-boson mass $\MZ = 91.1876$ GeV; the Fermi constant $G_F = 1.16638
\times 10^{-5}$ GeV$^{-2}$; the physical top and tau masses $M_t =
173.2\pm 0.9$ GeV and $M_\tau = 1.777$ GeV; and the running bottom mass
$m_b(m_b) = 4.18$ GeV. We use the one-loop formulae given in the
appendix A of ref.~\cite{Bernal:2007uv} to convert all the SM inputs
into $\msbar$ running parameters of the FSSM evaluated at the scale
$\MZ$. However, in view of the sensitivity of $\lambda$ to the precise
value of the top Yukawa coupling $g_t$, we include the two-loop QCD
contribution to the relation between the physical top mass $M_t$ and
its $\msbar$ counterpart $m_t$. In particular, we use:
\beq
m_t(\MZ) ~=~ \frac{M_t}{1+\frac{g_3^2}{(4\pi)^2}\,C_1
  ~+~\frac{g_3^4}{(4\pi)^4}\,
\left(C_2^\smallSM + C_2^{\tgp}\right)} ~+~ \Sigma_t(m_t)^{\rm EW}~,
\eeq
where $g_3$ is computed at the scale $\MZ$ using eq.~(A.1) of
ref.~\cite{Bernal:2007uv}, $\Sigma_t(m_t)^{\rm EW}$ denotes the terms
in the one-loop top self energy that do not involve the strong
interaction, and
\bea
C_1 &=& \frac{16}{3}-4\,\ln\frac{M_t^2}{\MZ^2}~,\\
\label{eq:C2SM}
C_2^{\,\smallSM} &=&\frac{2821}{18}+\frac{16}{3}\,\zeta_2\,(1+\ln4)
-\frac83\,\zeta_3
-\frac{338}{3}\,\ln\frac{M_t^2}{\MZ^2}+22\,\ln^2\frac{M_t^2}{\MZ^2}~,\\
\label{eq:C2g}
C_2^\tgp &=& \frac{89}{9} + 4\,\ln\frac{\mtgp^2}{\MZ^2}\,\left(
\frac{13}{3}+\ln\frac{\mtgp^2}{\MZ^2}-2\,\ln\frac{M_t^2}{\MZ^2}\right)~.
\eea
The boundary condition for the top Yukawa coupling of the FSSM is then
given by $g_t(\MZ)= m_t(\MZ)/v(\MZ)$. The two-loop SM contribution
$C_2^{\,\smallSM}$ in eq.~(\ref{eq:C2SM}) is from
ref.~\cite{Fleischer:1998dw}, while to obtain the two-loop F-gluino
contribution $C_2^\tgp$ in eq.~(\ref{eq:C2g}) we adapted the results
of ref.~\cite{Avdeev:1997sz} to the case of a heavy Majorana fermion
in the adjoint representation of $SU(3)$. For an F-gluino mass of a
few TeV, the inclusion of $C_2^\tgp$ in the boundary condition for
$g_t$ becomes crucial, as it changes the prediction for the Higgs mass
by several GeV. Alternatively, one could decouple the F-gluino
contribution from the RG evolution of the couplings below the scale
$\Mtgp$, include only the SM contributions in the boundary conditions
for $g_t$ and $g_3$ at the scale $\MZ$, and include the
non-logarithmic part of $C_2^\tgp$ as a threshold correction to $g_t$
at the scale $\Mtgp$. We have checked that the predictions for the
Higgs mass obtained with the two procedures are in very good agreement
with each other.

To improve our determination of the quartic coupling $\lambda$ at the
weak scale, we use two-loop renormalisation-group equations (RGEs) to
evolve the couplings of the effective theory between the scales $\MS$
and $\MZ$. Results for the two-loop RGEs of Split SUSY have been
presented earlier in
refs~\cite{Binger:2004nn,Giudice:2011cg,Tamarit:2012ie}. Since there
are discrepancies between the existing calculations, we used the
public codes {\tt SARAH}~\cite{Staub:2013tta} and {\tt
  PyR@TE}~\cite{Lyonnet:2013dna} to obtain independent results for the
RGEs of Split SUSY in the $\msbar$ scheme. Taking into account the
different conventions, we agree with the RGE for $\lambda$ presented
in ref.~\cite{Binger:2004nn}, and with all the RGEs for the
dimensionless couplings presented in section 3.1 of
ref.~\cite{Tamarit:2012ie}. However, we disagree with
ref.~\cite{Tamarit:2012ie} in some of the RGEs for the mass parameters
(our results for the latter are collected in the appendix).
Concerning the RGEs for the dimensionless couplings presented in
ref.~\cite{Giudice:2011cg}, we find some discrepancies~\footnote{~In
  particular, in ref.~\cite{Giudice:2011cg} the coefficient of $g_2^4$
  in the RGEs for $g_t,\,g_b,\,g_\tau,\,\tilde{g}_{1u}$ and
  $\tilde{g}_{1d}$ should be changed from $-15/4$ to $-17/4$, while
  the coefficient of $g_2^4$ in the RGEs for $\tilde{g}_{2u}$ and
  $\tilde{g}_{2d}$ should be changed from $-121/4$ to $-409/12$. In
  the RGE for $\lambda$, the terms proportional to $g_2^6,\,\lambda
  g_2^4$ and $g_2^4 g_1^2$ should be corrected in accordance with
  ref.~\cite{Binger:2004nn}. We thank A.~Strumia for confirming these
  corrections.} in two-loop terms proportional to $g_2^4$ and $g_2^6$.

At the end of our iterative procedure, we evolve all the parameters to
a common weak scale $Q_\smallW$, and obtain the physical squared mass
for the Higgs boson as
\bea
\label{eq:mH}
\MH^2 &=&\frac{\lambda(Q_\smallW)}{\sqrt2\,G_F}\,
\left[1-\delta^{1\ell}(Q_\smallW)\right]\nonumber\\
&+&\frac{g_t^4\,v^2}{128\,\pi^4}\,\left[
16\,g_3^2\,(3\,\ell_t^2+\ell_t)
-3\,g_t^2\,\left(9\,\ell_t^2-3\,\ell_t+2+\frac{\pi^2}3\right)\right]\nonumber\\
&+&\frac{g_3^4\,g_t^4\,v^2}{64\,\pi^6}\,\ln^3\frac{\mtgp^2}{Q_\smallW^2}~,
\eea
where $\ell_t = \ln(m_t^2/Q_\smallW^2)$. The one-loop correction
$\delta^{1\ell}(Q_\smallW)$, which must be computed in terms of
$\msbar$ parameters, is given in eqs~(15a)--(15f) of
ref.~\cite{Sirlin:1985ux}, while the two-loop corrections proportional
to $g_3^2g_t^4$ and to $g_t^6$ come from
ref.~\cite{Degrassi:2012ry}. We have also included the
leading-logarithmic correction arising from three-loop diagrams
involving F-gluinos, which can become relevant for large values of
$\mtgp/Q_\smallW$. This last term must of course be omitted if the
F-gluinos are decoupled from the RGE for $\lambda$ below the scale
$\Mtgp$. In our numerical calculations we set $Q_\smallW= M_t$ to
minimise the effect of the radiative corrections involving top quarks,
but we have found that our results for the physical Higgs mass are
remarkably stable with respect to variations of $Q_\smallW$.

\subsection{Results}
\label{sec:results}

\begin{figure}[t]
\begin{center}
\includegraphics[width=0.8\textwidth]{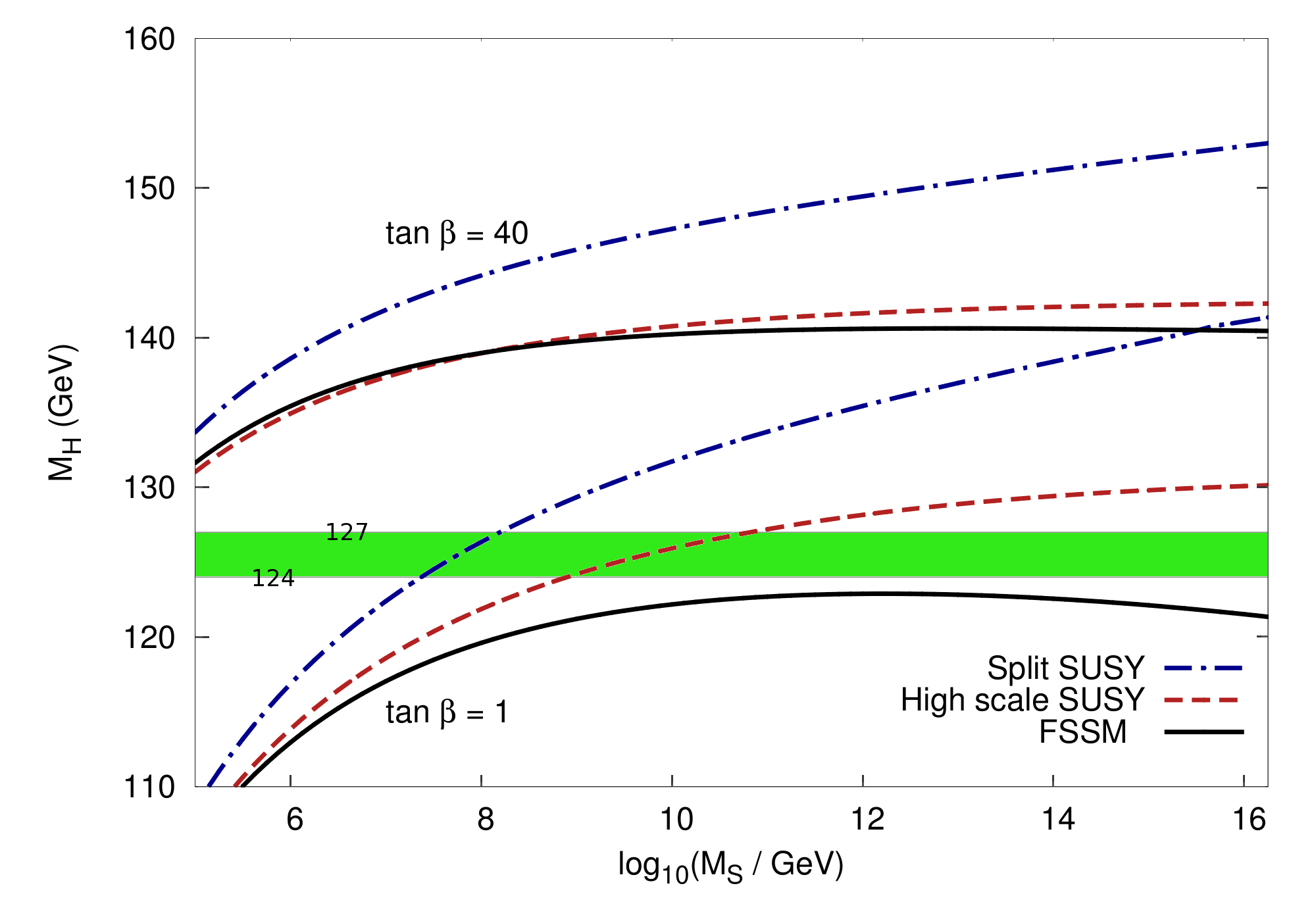}
\caption{ \footnotesize Higgs-mass predictions as a function of the
  SUSY scale $\MS$ for FSSM, High-Scale SUSY and Split SUSY.  We set
  $\Mtgp = \mu= 2$ TeV and $\tan\beta=1$ or $40$. The green-shaded
  region indicates a Higgs mass in the range $[124,127]$ GeV. }
\vspace*{-2mm}
\label{fig:mhiggs_multi}
\end{center}
\end{figure}

We find that, in the FSSM, the dependence of the physical Higgs mass
on the SUSY scale $\MS$ differs markedly from the cases of regular
Split SUSY or High-Scale SUSY (where all superparticle masses are set
to the scale $\MS$). Figure~\ref{fig:mhiggs_multi} illustrates this
discrepancy, showing $\MH$ as a function of $\MS$ for $\Mtgp=\mu=2$
TeV.  The solid (black) curves represent the prediction of the FSSM,
the dashed (red) ones represent the prediction of High-Scale SUSY, and
the dot-dashed (blue) ones represent the prediction of regular Split
SUSY (the predictions for the latter two models were obtained with
appropriate modifications of the FSSM calculation described in section
\ref{sec:procedure}). For each model, the lower curves were obtained
with $\tan\beta=1$, resulting in the lowest possible value of $\MH$
for a given $\MS$, while the upper curves were obtained with
$\tan\beta=40$.

As was shown earlier in ref.~\cite{Giudice:2011cg}, the Higgs mass
grows monotonically with the SUSY scale $\MS$ in the Split-SUSY case,
while it reaches a plateau in High-Scale SUSY. In both cases, the
prediction for the Higgs mass falls between $124$ and $127$ GeV only
for a relatively narrow range of $\MS$, well below the unification
scale $\MGUT \approx 2\!\times\!10^{16}$ GeV. In the FSSM, on the
other hand, the Higgs mass reaches a maximum and then starts
decreasing, remaining generally lower than in the other models. It is
therefore much easier to obtain a Higgs mass close to the
experimentally observed value even for large values of the SUSY scale.
For example, as will be discussed later, when $\tan\beta \approx 1.5$
we find that the FSSM prediction for the Higgs mass falls between
$124$ and $127$ GeV for all values of $\MS$ between $10^8$ GeV and
$\MGUT\,$.

\begin{figure}[t]
\begin{center}
\includegraphics[width=0.8\textwidth]{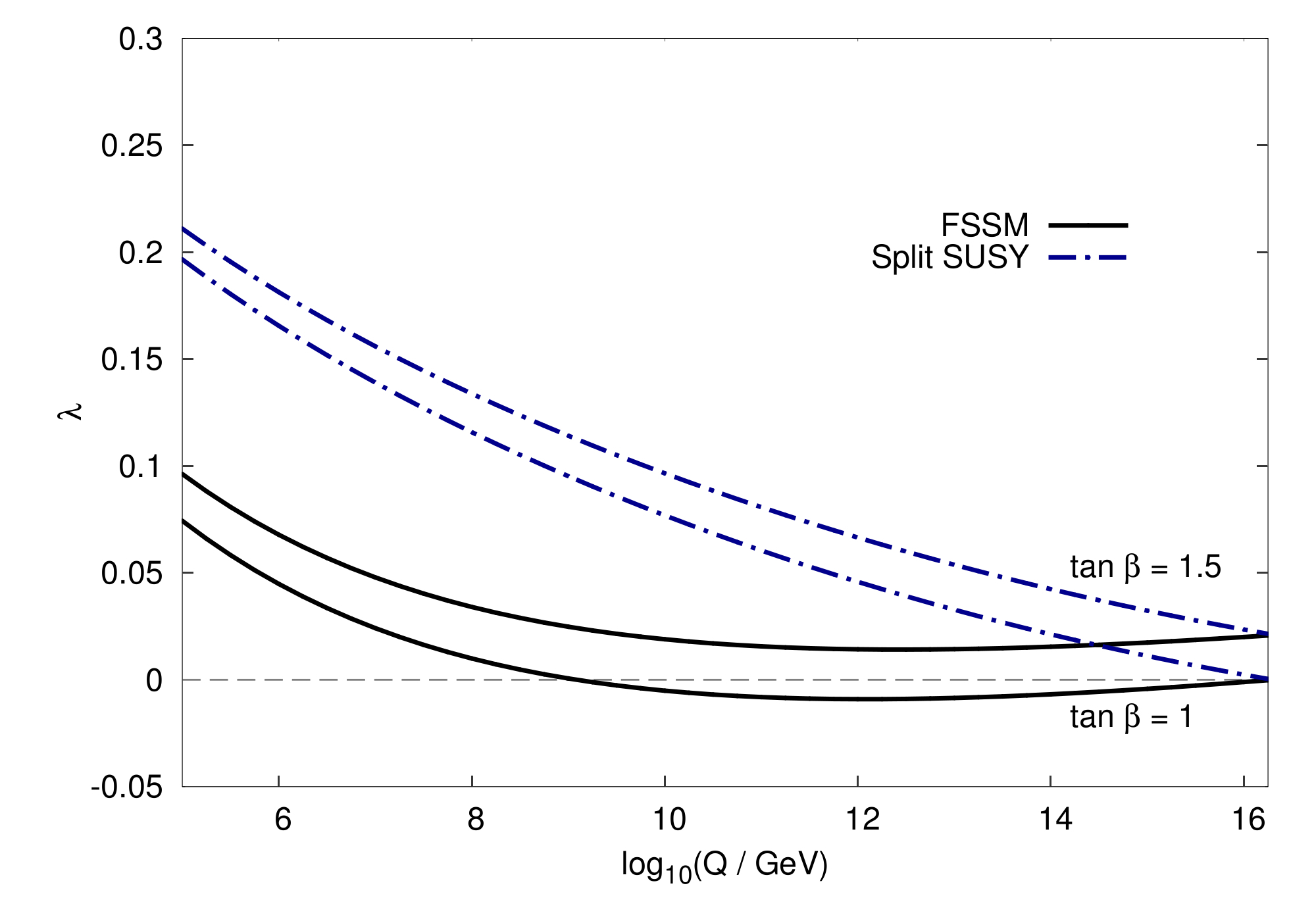}
\caption{\footnotesize Running of the Higgs quartic coupling $\lambda$
  in the FSSM and in the usual Split-SUSY case for $\tan\beta=1$ and
  $1.5$. We set $\MS = 2\!\times\! 10^{16}$ GeV and $\Mtgp= \mu = 2$
  TeV.}
\label{fig:lambda_run}
\vspace*{-2mm}
\end{center}
\end{figure}

This new behaviour originates in the RG evolution of $\lambda$ in the
FSSM, which differs from the case of Split SUSY. In figure
\ref{fig:lambda_run} we show the dependence of $\lambda$ on the
renormalisation scale $Q$ in the two theories, imposing the boundary
condition in eq.~(\ref{lambdaMS}) at the scale
$\MS=2\!\times\!10^{16}$ and setting $\tan\beta$ to either $1$ or
$1.5$. Even though we impose the same boundary condition in both
theories, the fact that the effective Higgs--higgsino--gaugino couplings
are zero in the FSSM induces a different evolution. Indeed, in Split
SUSY the contributions proportional to four powers of the
Higgs--higgsino--gaugino couplings enter the one-loop part of
$\beta_{\lambda}$ with negative sign, as do those proportional to four
powers of the top Yukawa coupling, whereas the contributions
proportional to four powers of the gauge couplings enter with positive
sign. For $\MS\gtrsim 10^{12}$ GeV, the top Yukawa coupling is
sufficiently suppressed at the matching scale that removing the
Higgs--higgsino--gaugino couplings makes $\beta_{\lambda}$
positive. This prompts $\lambda$ to decrease with decreasing $Q$,
until the negative contribution of the top Yukawa coupling takes over
and $\lambda$ begins to increase.

\begin{figure}[p]
\begin{center}
\includegraphics[height=0.4\textheight]{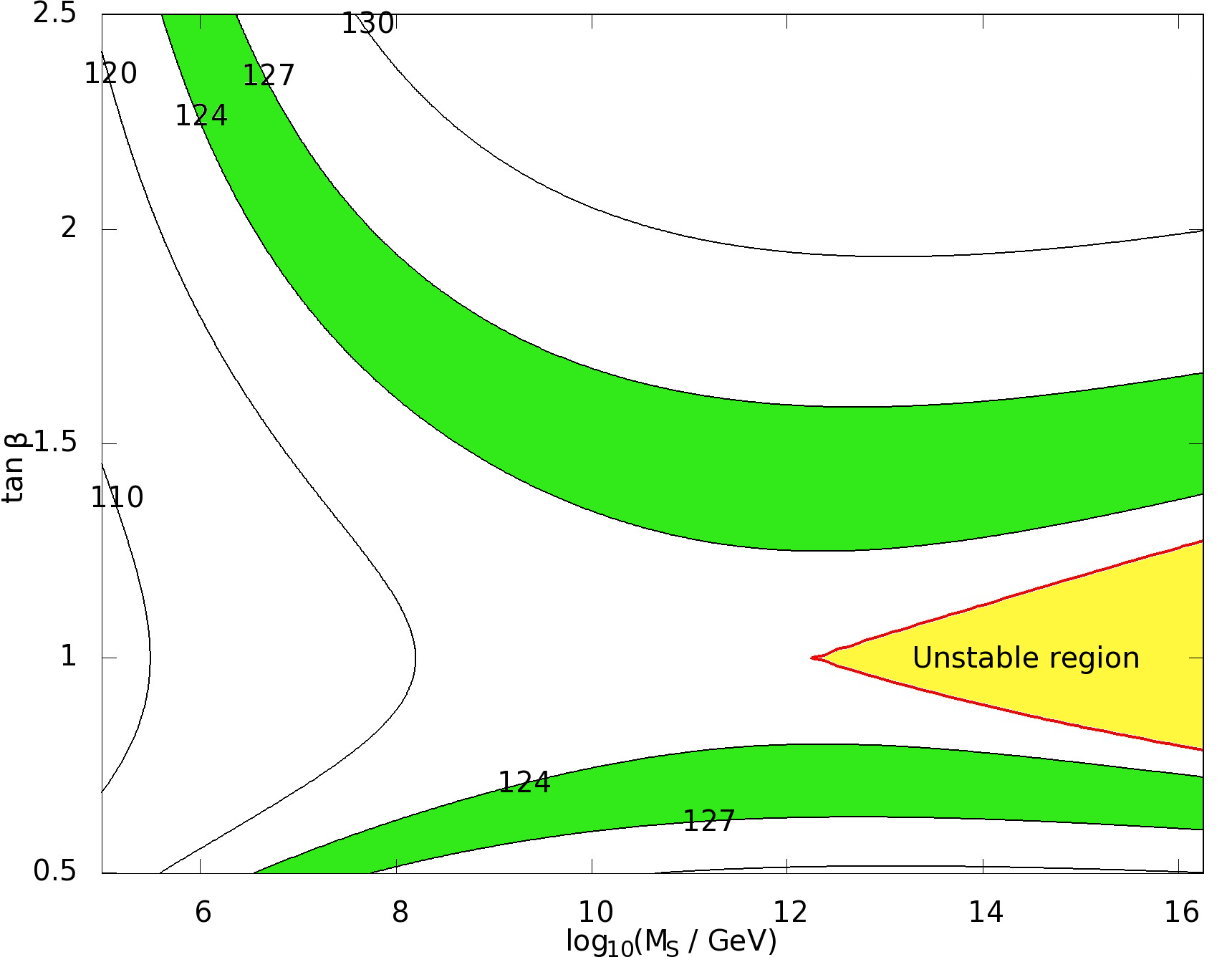}
\caption{ \footnotesize Contour plot of the prediction for the Higgs
  mass on the $\MS\,$--$\,\tan\beta$ plane, for $\Mtgp = \mu= 2$
  TeV. The yellow-shaded region indicates where $\lambda$ becomes
  negative during its running between $\MZ$ and $\MS$. The
  green-shaded region indicates a Higgs mass in the range $[124,127]$
  GeV.}
\label{fig:mh_tb_ms}
\end{center}
\end{figure}

Figure \ref{fig:lambda_run} also shows that, for values of $\tan\beta$
sufficiently close to 1, the quartic coupling $\lambda$ can become
negative during its evolution down from the scale $\MS$, only to
become positive again when $Q$ approaches the weak scale. This points
to an unstable vacuum, and a situation similar to the one described in
ref.~\cite{Butazzo:2013}. However, it was already clear from
figure~\ref{fig:mhiggs_multi} that, for $\tan\beta=1$, the FSSM
prediction for the Higgs mass is too low anyway. For the values of
$\tan\beta$ large enough to induce a Higgs mass in the observed range,
the theory is stable. This is illustrated in
figure~\ref{fig:mh_tb_ms}, where we show the contours of equal Higgs
mass on the $\MS\,$--$\,\tan\beta$ plane, setting $\Mtgp = \mu= 2$
TeV. The green-shaded region corresponds to a Higgs mass in the
observed range between $124$ and $127$ GeV, while the yellow-shaded
region is where $\lambda$ becomes negative during its evolution
between $\MS$ and the weak scale, and the vacuum is unstable.
It can be seen that, for $\MS \gtrsim 10^8$ GeV, a Higgs mass around
$126$ GeV can be comfortably obtained for either $\tan\beta \approx
1.5$ or $\tan\beta \approx 0.6$. The unstable region is confined to
values of $\tan\beta$ very close to $1$, and only for $\MS \gtrsim
10^{12}$ GeV.  For lower values of $\MS$, the top Yukawa coupling is
not sufficiently suppressed at the matching scale and
$\beta_{\lambda}$ is always negative, therefore there is no region of
instability.

We investigated how our results are affected by the experimental
uncertainty on the top mass.  An increase (or decrease) of $1$ GeV
from the central value $M_t=173.2$ GeV used in
figure~\ref{fig:mh_tb_ms} translates into an increase (or decrease) in
our prediction for the Higgs mass of $1$--$2$ GeV, depending on
$\MS$. For larger values of $M_t$, the observed value of $\MH$ is
obtained for $\tan\beta$ closer to $1$, and the green regions in
figure~\ref{fig:mh_tb_ms} approach the unstable region. The size of
the unstable region is itself dependent on $M_t$ (i.e.~the region
shrinks for larger $M_t$) but the effect is much less
pronounced. Consequently, raising the value of the top mass may lead
to instability for large $\MS$ (e.g.~for $\MS \gtrsim 10^{12}$ GeV
when $\Mtgp = 2$ TeV).  Considering an extreme case, for $M_t = 175$
GeV we would see a substantial overlap of the experimentally
acceptable regions with the unstable region around $\MS \approx
\MGUT$.  On the other hand, for values of $M_t$ lower than $173.2$ GeV
the green regions in figure~\ref{fig:mh_tb_ms} are shifted towards
values of $\tan\beta$ further away from $1$, and the vacuum is always
stable for the correct Higgs mass.

\begin{figure}[p]
\begin{center}
\includegraphics[height=0.4\textheight]{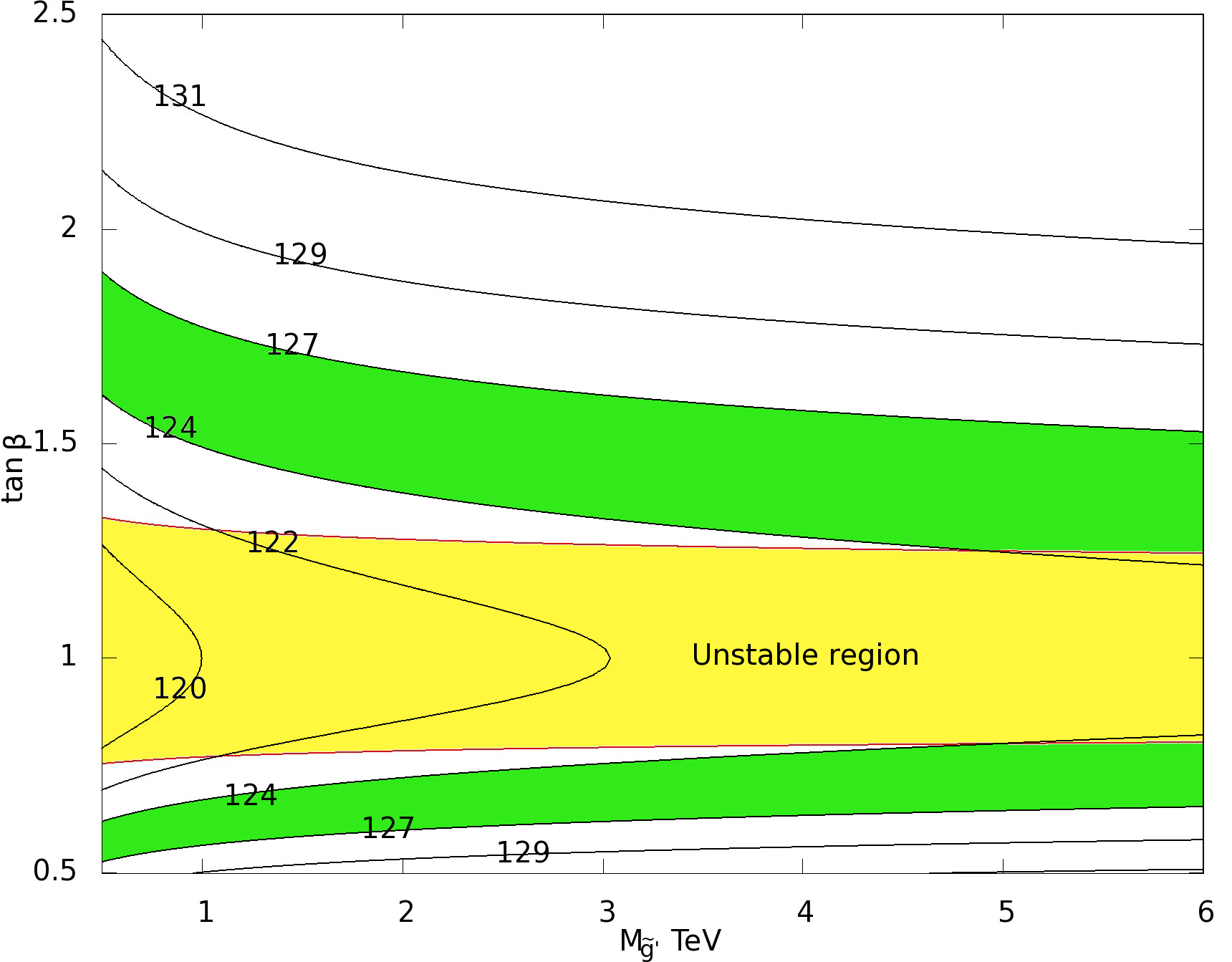}
\caption{ \footnotesize Same as figure \ref{fig:mh_tb_ms} on the
  $\Mtgp\,$--$\,\tan\beta$ plane, with $\MS = 2 \!\times\!  10^{16}$
  GeV and $\mu = 2$ TeV. }
\label{fig:mh_mg_tanb}
\end{center}
\end{figure}

Finally, in figure \ref{fig:mh_mg_tanb} we show the contours of equal
Higgs mass on the $\Mtgp\,$--$\,\tan\beta$ plane, setting $\MS = 2
\!\times\!  10^{16}$ GeV and $\mu = 2$ TeV. The colour code is the
same as in figure~\ref{fig:mh_tb_ms}. It can be seen that the region
where the FSSM prediction for the Higgs mass is between $124$ and
$127$ GeV gets closer to the unstable region when the F-gluino mass
increases.  However, the dependence of $\MH$ on $\Mtgp$ is relatively
mild, and only when $\Mtgp$ is in the multi-TeV region do the green
and yellow regions in figure \ref{fig:mh_mg_tanb} overlap.
We conclude that if we insist on enforcing exact stability and setting
$\MS \approx 2 \!\times\!  10^{16}$ GeV, then obtaining a Higgs mass
compatible with the observed value constrains the gluino mass to the
few-TeV region.

\section{Conclusions}
\label{SEC:Conclusions}

We have defined a model -- the FSSM -- which has the same particle content
at low energies as Split SUSY, but has a substantially different
ultraviolet completion and also low-energy phenomenology:

\begin{enumerate}

\item We discussed in section \ref{sec:fssm} that the F-gaugino and
  F-higgsino couplings to the Higgs are suppressed by $\varepsilon^2$.

\item The effective operators leading to the decay of the
  charginos/heavier neutralinos, which are generated by integrating
  out the sfermions, are also suppressed, because the adjoint fermions
  $\chi$ do not have a gauge-current coupling to the sfermions.  As
  discussed in section \ref{SEC:Constraints}, the lifetimes are
  enhanced by a factor $\varepsilon^{-4}$. This makes the
  gauginos/higgsinos very long-lived; we must appeal to a non-thermal
  history of the universe with a low reheating temperature to avoid
  unwanted relics.

\item Since we no longer have an $R$-symmetry, the usual corrections to 
the Higgs quartic coupling at
  the SUSY scale proportional to powers of $|A_t- \mu_0 \cot \beta|^2/\MS^2$ are in
  principle no longer negligible. However, as we discussed in section
  \ref{SEC:Higgs}, in Split-SUSY scenarios those corrections are less
  important than in the MSSM, because the evolution to the large scale
  $\MS$ suppresses the top Yukawa coupling that multiplies
  them~\cite{Bernal:2007uv,Giudice:2011cg}.

\item Finally, the main result of this paper was presented in section
  \ref{SEC:Higgs}, and concerns the precision determination of the
  Higgs mass in this model. Its value is substantially different than
  in either High-Scale or Split SUSY; in particular we can find
  $126$ GeV for \emph{any} SUSY scale, with a vacuum that is always
  stable when the F-gluino mass is not too large.
\end{enumerate}

We have found that a standard-model-like Higgs boson with a mass
around $126$ GeV can be obtained for low values of $\tan \beta$. For
low values of $\MS$, the exact value of $\tan \beta$ is subject to
modification that we estimated when considering the presence of
additional contributions to the quartic Higgs coupling from the
unsuppressed $A$-terms. For larger values of $\MS$, the latter
contributions are negligible.

In supersymmetric theories, the theorem of non-renormalisation of the
superpotential implies that supersymmetry cannot be broken by
perturbative effects. It is either broken at tree level or by
non-perturbative effects.  The former implies that the scale of
supersymmetry breaking is of the order of the fundamental (string)
scale $M_ *$, and unless this is taken to lie at an intermediate
energy scale~\cite{Benakli:1998pw}, it predicts a heavy spectrum.  In
studies of low-energy supersymmetry, the use of non-perturbative
effects attracted most interest because it allows the generation of
the required large hierarchy of scales through dimensional
transmutation.  It is then interesting to investigate the fate of the
former possibility when the supersymmetry scale is pushed to higher
values.  For Split and High-Scale SUSY, it is difficult to justify a
very high $\mathcal{O}(\MGUT)$ SUSY scale, since in that regime they
predict the Higgs mass to be too high (unless one pushes to the limits
of the theoretical and experimental uncertainties, see
e.g.~refs\cite{Giudice:2011cg,Delgado:2013gza}).

Here, we have shown that the situation is different in the Fake Split
SUSY Model.  It is tempting to consider that while supersymmetry is
broken at tree level in a secluded sector, the scale $\MS \sim \MGUT$
could be induced through radiative effects \cite{Benakli:2010gi} from
the fundamental scale $\MS \sim \alpha M_*\,$, where $\alpha$ is a
loop factor. We postpone the construction of explicit realisations of
this possibility for a future study.

\section*{Acknowledgments}
We thank Emilian Dudas, Jose Ramon Espinosa, Mariano Quir\'os, Alessandro Strumia and Carlos Tamarit for useful discussions.

\newpage


\section*{Appendix:  Two-loop RGEs for Split-SUSY masses}
\addcontentsline{toc}{section}{Appendix:  Two-loop RGEs for Split-SUSY masses}

\begin{Appendix}

In this appendix we list the two-loop RGEs for the fermion-mass parameters
of Split SUSY in the $\msbar$ scheme. 
Defining
\begin{align}
\frac{dm_x}{d\ln Q} = \frac{\beta^{(1)}_{m_x}}{16 \pi^2} ~+~
\frac{\beta^{(2)}_{m_x}}{256 \pi^4}
\,,~~~~~~~~(m_x = m_{\tilde g},\,\mtB,\,\mtW,\,\mu)~,
\end{align}
we obtained, using the public codes {\tt SARAH}~\cite{Staub:2013tta} and {\tt
  PyR@TE}~\cite{Lyonnet:2013dna},
\bea 
\beta_{m_{\tilde{g}} }^{(1)}  &=  &
-18 g_{3}^{2} m_{\tilde{g}} ~,~~~~~~~~~~ 
\beta_{m_{\tilde{g}} }^{(2)}  ~=~
-228 g_{3}^{4} m_{\tilde{g}} ~, \\ 
\nonumber \\
\beta_{\mtB}^{(1)} &=  & ( \tilde{g}_{1\mathrm{u}}^{2}  
+ \tilde{g}_{1\mathrm{d}}^{2} ) \mtB +
4 \tilde{g}_{1\mathrm{d}}  \tilde{g}_{1\mathrm{u}}  \mu  ~,  \\ 
\beta_{\mtB}^{(2)}  &=  &
\left[ \frac{1}{8} \left(\tilde{g}_{1\mathrm{u}}^{4}  
+ \tilde{g}_{1\mathrm{d}}^{4}\right) 
-\frac{7}{2} \tilde{g}_{1\mathrm{d}}^{2} \tilde{g}_{1\mathrm{u}}^{2} 
-\frac{21}{8} \left(\tilde{g}_{1\mathrm{u}}^{2} \tilde{g}_{2\mathrm{u}}^{2}   
+\tilde{g}_{1\mathrm{d}}^{2} \tilde{g}_{2\mathrm{d}}^{2}\right)        
-\frac{9}{4} \left(\tilde{g}_{1\mathrm{u}}^{2} \tilde{g}_{2\mathrm{d}}^{2}  
+\tilde{g}_{1\mathrm{d}}^{2} \tilde{g}_{2\mathrm{u}}^{2}\right)
 \right. \nonumber \\ 
 &&\left. \, \, \, \,      
+ \frac{51}{8} \,\left(\tilde{g}_{1\mathrm{u}}^{2} 
+\tilde{g}_{1\mathrm{d}}^{2}\right) 
\left(  \frac{1}{5} g_{1}^{2} + g_{2}^{2}  \right)
-\frac32\left( {\tilde{g}_{1\mathrm{u}}^{2} 
+\tilde{g}_{1\mathrm{d}}^{2} }\right) \left(3g_{b }^{2} 
+ 3g_{t }^{2}+ g_{\tau }^{2} \right) \right] 
\mtB \nonumber \\ 
 & +&\left[ 3 \tilde{g}_{2\mathrm{d}}^{2} \tilde{g}_{1\mathrm{d}}^{2} 
+3 \tilde{g}_{2\mathrm{u}}^{2} \tilde{g}_{1\mathrm{u}}^{2}  
- 6 \tilde{g}_{2\mathrm{d}}  \tilde{g}_{1\mathrm{d}}  
\tilde{g}_{2\mathrm{u}}  
\tilde{g}_{1\mathrm{u}}   \right]  \mtW  \nonumber \\ 
 & +&\left[ \frac{24}{5} g_{1}^{2}   +24 g_{2}^{2}  
- \tilde{g}_{1\mathrm{u}}^{2}- \tilde{g}_{1\mathrm{d}}^{2}   
 -3 \tilde{g}_{2\mathrm{u}}^{2}-3 \tilde{g}_{2\mathrm{d}}^{2}    
\right]\,  \tilde{g}_{1\mathrm{d}}  \tilde{g}_{1\mathrm{u}} \mu  ~, \\ 
\nonumber \\ 
\beta_{\mtW }^{(1)}  &=  &
\left(-12 g_{2}^{2}       + \tilde{g}_{2\mathrm{u}}^{2} 
+ \tilde{g}_{2\mathrm{d}}^{2} \right) \mtW  
+ 4 \tilde{g}_{2\mathrm{d}} \tilde{g}_{2\mathrm{u}}   \mu ~, \\ 
\beta_{\mtW }^{(2)}  &= &  \left[
-\frac{29}{8} \left(\tilde{g}_{2\mathrm{u}}^{4} 
+\tilde{g}_{2\mathrm{d}}^{4}\right) 
-\frac{21}{2} \tilde{g}_{2\mathrm{d}}^{2} \tilde{g}_{2\mathrm{u}}^{2}  
-\frac{7}{8} \left(\tilde{g}_{2\mathrm{u}}^{2} \tilde{g}_{1\mathrm{u}}^{2}   
+\tilde{g}_{2\mathrm{d}}^{2} \tilde{g}_{1\mathrm{d}}^{2}\right)
-\frac{3}{4} \left(\tilde{g}_{2\mathrm{u}}^{2} \tilde{g}_{1\mathrm{d}}^{2}  
+\tilde{g}_{2\mathrm{d}}^{2} \tilde{g}_{1\mathrm{u}}^{2}\right)  
\right. \nonumber \\ 
 &&\left.  \, \, \, \,
-\frac{233}{3} g_{2}^{4} +\frac18 \left( \tilde{g}_{2\mathrm{u}}^{2} 
+\tilde{g}_{2\mathrm{d}}^{2} \right) \left(  \frac{51}{5} 
g_{1}^{2} +{91} g_{2}^{2}  \right)
-\frac{3}{2}\left( \tilde{g}_{2\mathrm{u}}^{2} 
+ \tilde{g}_{2\mathrm{d}}^{2} \right) \left(3 g_{b }^{2} 
+3 g_{t }^{2} + g_{\tau }^{2}\right) \right] 
\mtW  \nonumber \\ 
 &+& \left[\tilde{g}_{2\mathrm{u}}^{2} \tilde{g}_{1\mathrm{u}}^{2} +
\tilde{g}_{2\mathrm{d}}^{2} \tilde{g}_{1\mathrm{d}}^{2}  
-2  \tilde{g}_{2\mathrm{u}}  \tilde{g}_{1\mathrm{u}}  
\tilde{g}_{2\mathrm{d}}  \tilde{g}_{1\mathrm{d}}    
\right] \mtB  \nonumber \\ 
& +&\left[ \frac{24}{5} g_{1}^{2} +48 g_{2}^{2}     
- \tilde{g}_{1\mathrm{u}}^{2} - \tilde{g}_{1\mathrm{d}}^{2} 
-3 \tilde{g}_{2\mathrm{u}}^{2} -3 \tilde{g}_{2\mathrm{d}}^{2}
\right]\, \tilde{g}_{2\mathrm{d}} \tilde{g}_{2\mathrm{u}}\mu ~, \\ 
\nonumber \\ 
\beta_{\mu }^{(1)} &=  & \left(  -\frac{9}{2} g_{2}^{2} 
-\frac{9}{10} g_{1}^{2}  +\frac{3}{4} \tilde{g}_{2\mathrm{u}}^{2}  
+\frac{3}{4} \tilde{g}_{2\mathrm{d}}^{2}  
+\frac{1}{4} \tilde{g}_{1\mathrm{u}}^{2}  
+\frac{1}{4} \tilde{g}_{1\mathrm{d}}^{2}       \right) \mu  
+3 \tilde{g}_{2\mathrm{u}} \tilde{g}_{2\mathrm{d}}   \mtW 
+ \tilde{g}_{1\mathrm{u}} \tilde{g}_{1\mathrm{d}}   \mtB ~,\nonumber \\   \\ 
\beta_{\mu }^{(2)}  &= & \left[ -\frac{421}{16} g_{2}^{4} +
 \frac{1359}{400} g_{1}^{4}  -\frac{27}{40} g_{1}^{2} g_{2}^{2}  
-\frac{15}{8} \tilde{g}_{2\mathrm{u}}^{4}  
- \frac{15}{8} \tilde{g}_{2\mathrm{d}}^{4} 
-\frac{1}{4} \tilde{g}_{1\mathrm{u}}^{4}   
-\frac{1}{4} \tilde{g}_{1\mathrm{d}}^{4}  \right. \nonumber \\ 
 &&   \, \, \, \,+\frac{33}{160} g_{1}^{2} \left( \tilde{g}_{1\mathrm{u}}^{2} 
+ \tilde{g}_{1\mathrm{d}}^{2} +3 \tilde{g}_{2\mathrm{u}}^{2} 
+3 \tilde{g}_{2\mathrm{d}}^{2}\right)   
+\frac{33}{32} g_{2}^{2} \left( \tilde{g}_{1\mathrm{u}}^{2}
+ \tilde{g}_{1\mathrm{d}}^{2}  +11 \tilde{g}_{2\mathrm{d}}^{2} 
+ 11 \tilde{g}_{2\mathrm{u}}^{2}  \right)   \nonumber \\ 
 &&   \, \, \, \, -\frac{9}{8} \left(
\tilde{g}_{2\mathrm{u}}^{2} \tilde{g}_{1\mathrm{u}}^{2}   
+\tilde{g}_{2\mathrm{u}}^{2} \tilde{g}_{1\mathrm{d}}^{2}     
+\tilde{g}_{2\mathrm{d}}^{2} \tilde{g}_{1\mathrm{u}}^{2}  
+\tilde{g}_{2\mathrm{d}}^{2} \tilde{g}_{1\mathrm{d}}^{2}\right)   
-\frac{45}{4} \tilde{g}_{2\mathrm{d}}^{2} \tilde{g}_{2\mathrm{u}}^{2}   
-2 \tilde{g}_{1\mathrm{d}}^{2} \tilde{g}_{1\mathrm{u}}^{2}       
\nonumber \\ 
 &&\left. \, \, \, \,+3 \tilde{g}_{2\mathrm{d}}  \tilde{g}_{1\mathrm{d}}  
\tilde{g}_{2\mathrm{u}}  \tilde{g}_{1\mathrm{u}}      
-\frac{3}{8} \left(
\tilde{g}_{1\mathrm{u}}^{2} + \tilde{g}_{1\mathrm{d}}^{2} +
3 \tilde{g}_{2\mathrm{u}}^{2} + 3\tilde{g}_{2\mathrm{d}}^{2} \right) 
\left( 3g_{t }^{2} + 3g_{b }^{2}  +  g_{\tau }^{2} \right)  
\right] \mu \nonumber \\ 
 &+&\left[ \frac{87}{2} g_{2}^{2}   
+ \frac{27}{10} g_{1}^{2}   
-3 \tilde{g}_{2\mathrm{u}}^{2} -3 \tilde{g}_{2\mathrm{d}}^{2} 
\right] \tilde{g}_{2\mathrm{u}} \tilde{g}_{2\mathrm{d}}
\mtW +\left[ \frac{9}{2} g_{2}^{2} + \frac{9}{10} g_{1}^{2} 
- \tilde{g}_{1\mathrm{u}}^{2} - \tilde{g}_{1\mathrm{d}}^{2} \right] 
 \tilde{g}_{1\mathrm{u}} \tilde{g}_{1\mathrm{d}} \mtB \,. \nonumber \\ 
\eea
\end{Appendix}


\section*{Note Added}
After the appearance of our paper in preprint, the author of ref.~\cite{Tamarit:2012ie} revised his calculation of the two-loop RGEs in Split SUSY. His results for the RGEs of the fermion-mass parameters are now in full agreement with ours.

\end{document}